\documentclass[a4paper]{PoS}
\usepackage{amsmath,amssymb}
\usepackage{array,dsfont}
\usepackage{float}
\usepackage[justification=justified,singlelinecheck=false]{caption,subfig}
\usepackage{placeins}
\newcolumntype{M}[1]{>{\centering\arraybackslash}m{#1}}
\newcolumntype{N}{@{}m{0pt}@{}}

\newcommand{\LL}{\mathcal{L}}
\newcommand{\tr}[1]{\mathop{\rm tr} \left[ #1\right]}

\newcommand{\ii}{{\rm i}}

\newcommand{\vV}{\mathbf{V}}

\newcommand{\vH}{\mathbf{H}}

\newcommand{\vD}{\mathbf{D}}

\newcommand{\bss}{\begin{tiny}}
\newcommand{\ess}{\end{tiny}}

\title{Effective Field Theory and Unitarity in Vector Boson Scattering}

\ShortTitle{Effective Field Theory and Unitarity in Vector Boson Scattering}

\author{
  \speaker{Marco Sekulla}$^a$,
  Wolfgang Kilian$^b$,
  Thorsten Ohl$^c$,
  J\"urgen Reuter$^d$ \\
  \llap{$ˆa$} Karlsruhe Institute of Technology (DE)\\
  \llap{$ˆb$} University of Siegen (DE)\\
  \llap{$ˆc$} W\"urzburg University (DE)\\
  \llap{$ˆd$} DESY Theory Group (DE)\\
  E-mail:
  \email{marco.sekulla@kit.edu},
  \email{kilian@physik.uni-siegen.de},
  \email{ohl@physik.uni-wuerzburg.de},
  \email{juergen.reuter@desy.de}}

\abstract{
Weak vector boson scattering at high energies will be one of the key measurements
in current and upcoming LHC runs. It is most sensitive to any new physics associated
with electroweak symmetry breaking. However, a conventional EFT analysis will fail at
high energies.
To address this problem, we present a parameter-free prescription valid for arbitrary
perturbative and non-perturbative models: the T-matrix unitarization. We describe
its implementation as an asymptotically consistent reference model matched to the
low-energy effective theory.
We show examples of typical observables of vector-boson scattering at the LHC in
our unitarized framework.
For many strongly-coupled models like composite Higgs models, dimension-8
operators might be actually the leading operators. In addition to those longitudinal
and transversal dimension eight EFT operators, the effects of generic tensor and scalar
resonances within simplified models are considered.
}

\FullConference{Fourth Annual Large Hadron Collider Physics\\
         13-18 June 2016\\
         Lund, Sweden}

\begin{document}

  \section{Motivation}
  Run I of the LHC has not only revealed a Standard Model-like Higgs
  boson~\cite{Aad:2012tfa,Chatrchyan:2012xdj} together with measuring
  its mass and some of its properties and couplings,
  but also established the scattering process of electroweak gauge
  bosons\cite{Aad:2014zda,ATLAS:2014rwa,CMS:2015jaa} (VBS) as
  predicted by the Standard Model (SM). This process gives insights into
  the nature of the electroweak symmetry breaking (EWSB) sector and
  further fundamental properties of the Higgs. In the SM, the
  electroweak breaking sector is described as a weakly interacting
  theory, where the Higgs boson is vastly suppressing the vector
  boson scattering process at high center-of-mass energies and the
  scattering amplitude is dominated by the transversal vector
  boson scattering.

  Without the Higgs the VBS scattering amplitudes $VV \rightarrow VV$,
  where $V$ is $W^\pm,Z$, would rise with $s / v^2$ due to the
  dominant contribution of scalar Goldstone-boson scattering,
  which represents the longitudinal degrees of freedom of the vector
  boson scattering. The electroweak interactions would become strongly
  interacting in the TeV range. However, the initial limits on VBS are
  rather weak and only scales close to the pair-production threshold
  of $\sim 200$ GeV are probed. Run II and III of the LHC and future
  (high-energy) $e^+e^-$ colliders will improve the accuracy
  and provide new insights in the origin of EWSB. The delicate
  cancellation between the EW gauge bosons and the Higgs boson in VBS
  makes this channel an ideal, yet intricate channel to search for new
  physics.

  The discussion in these proceedings is based on our publications
  in~\cite{Alboteanu:2008my,Kilian:2014zja,Kilian:2015opv,Sekulla:2015} and
  is an update of~\cite{Reuter:2016kqo}.


  \section{Effective Field Theory, Perturbative Unitarity and Unitarization}

  To study new physics in the VBS process generically, we
  will use the framework of Effective Field Theories (EFT). A set of
  higher-dimensional operators extends the SM Lagrangian to quantify
  deviations from the SM, which originate from some new physics at a
  high energy scale $\Lambda_i$ as
  \begin{align}
    \LL= \LL_{SM} + \sum_i \frac{C_i}{\Lambda_i^{d-4}} \mathcal{O}^{d}_i.
  \end{align}
  Here, $C_i$ are the associated Wilson coefficients of the operators.
  Lacking a possibility to disentangle both parameters, we introduce the ratio
  coupling $F_i=\frac{C_i}{\Lambda_i^{d-4}}$.

  Many different operator bases have been proposed for the electroweak
  sector, an overview and also translations between them have been
  discussed e.g. in~\cite{Baak:2013fwa,Degrande:2013rea}. For
  illustrative purposes, we study only the subset of operators which purely
  contribute to the Higgs sector and therefore affect only the coupling of longitudinal
  vector bosons and the Higgs boson. This subset of operators contains
  $\mathcal{O}_{HD}$ as a dim-6 operator, and the two dim-8 operators,
  $\mathcal{O}_{S,0}$ and $\mathcal{O}_{S,1}$ (cf. eqs. \eqref{LL-HD}--\eqref{LL-S1}).
  All of these operators
  could arise easily in popular scenarios of new physics beyond the SM
  (BSM) like Composite Higgs, Little Higgs or Extra Dimensions. The LHC
  experiments are studying all three of them to gain sensitivity in
  various channels like dibosons, tribosons, precision Higgs data and
  VBS. The operators are given by
  \begin{alignat}{3}
    \label{LL-HD}
    \LL_{HD} &=
    & & F_{HD}\ &&
    \tr{{\vH^\dagger\vH}- \frac{v^2}{4}}\cdot
    \tr{\left (\vD_\mu \vH \right )^\dagger \left (\vD^\mu \vH \right )} \, ,
    \\
    \label{LL-S0}
    \LL_{S,0}&=
    & &F_{S,0}\ &&
    \tr{ \left ( \vD_\mu \vH \right )^\dagger \vD_\nu \vH}
    \cdot \tr{ \left ( \vD^\mu \vH \right )^\dagger \vD^\nu \vH} \, ,
    \\
    \label{LL-S1}
    \LL_{S,1}&=
    & &F_{S,1}\ &&
    \tr{ \left ( \vD_\mu \vH \right )^\dagger \vD^\mu \vH}
    \cdot \tr{ \left ( \vD_\nu \vH \right )^\dagger \vD^\nu \vH} \, .
  \end{alignat}

  Due to the unknown microscopic picture of the underlying energy giving
  rise to these operators, the validity range of the EFT is also a priori
  unknown. In this case, the unitarity condition is used to determine the
  validity of the EFT.

  \begin{figure}[tb]
    \includegraphics[width=0.5\linewidth]{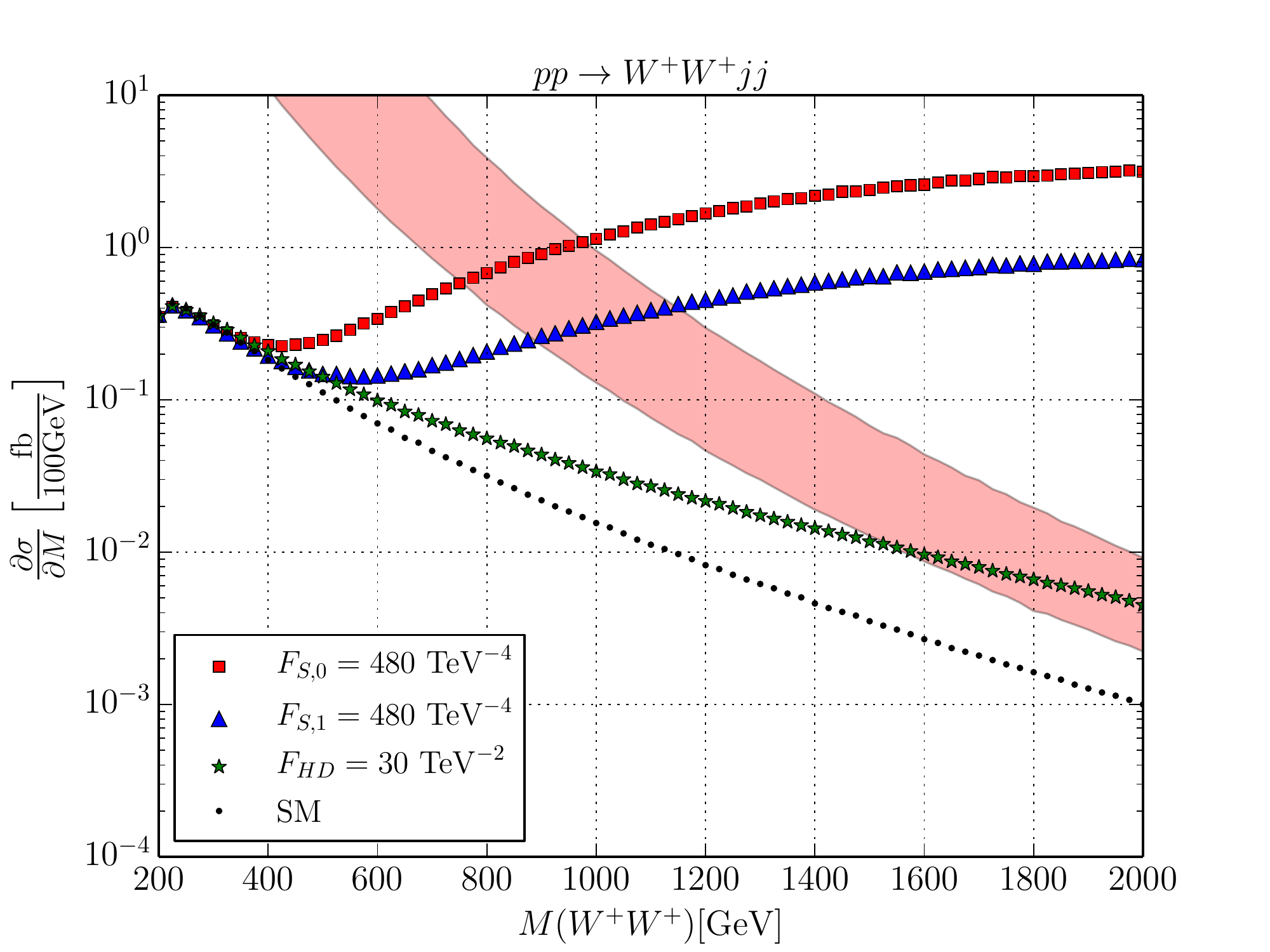}
    \includegraphics[width=0.5\linewidth]{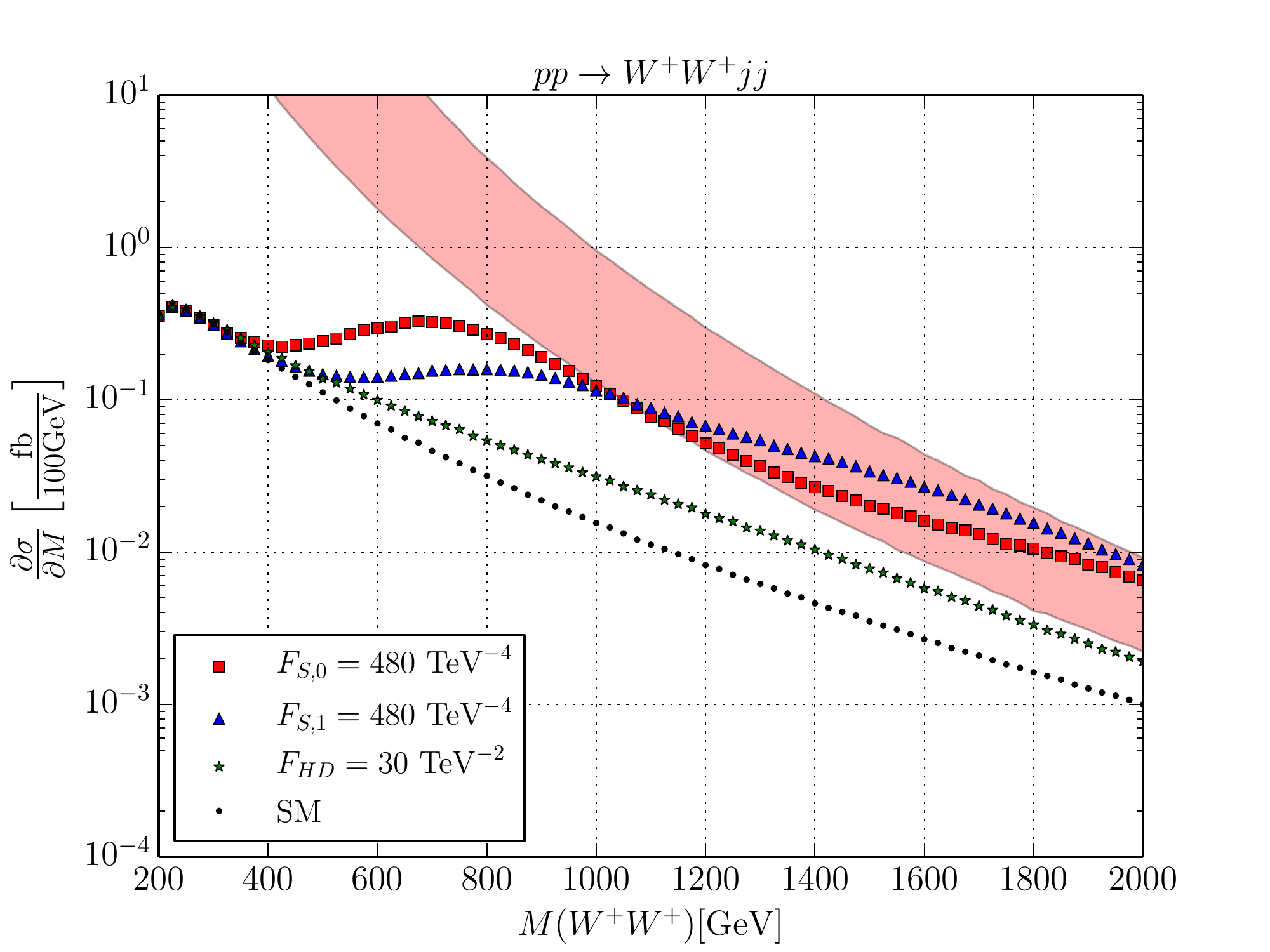}
    \caption{\label{fig:ww_unit}
      $pp\to W^+W^+jj$, left: naive EFT results that violate
      unitarity, QCD contributions neglected. The band describes maximal
      allowed values, due to unitarity constraints, for the
      differential cross section. The lower bound describes the
      saturation of isospin partial wave$ \mathcal{A}_{20}$
      and the upper bound describes the
      simultaneous saturation of $\mathcal{A}_{20}$ and
      $\mathcal{A}_{22}$, right : unitarized result.
      Cuts: $M_{jj} > 500$ GeV;
      $\Delta\eta_{jj} > 2.4$;
      $p^j_T > 20$ GeV;
      $|\eta_j| > 4.5$.}
  \end{figure}

  In the left-hand side of Fig.~\ref{fig:ww_unit}, the cross section for
  the complete LHC process $pp\rightarrow W^+W^+jj$ at leading order --
  computed using the Monte-Carlo generator \texttt{WHIZARD}~\cite{Kilian:2007gr,Moretti:2001zz}
  with CTEQ6L PDF sets -- is shown. The SM curve is compared to three curves
  for models which contain a single nonzero coefficient for the
  three different effective higher-dimensional operators, respectively.
  For an indication of the unitarity limits, we have included a
  quartic Goldstone interaction amplitude with a constant coefficient
  $a_{IJ}=\ii$ in the $I=2$ and $J=0,2$ isospin and spin channels and
  recomputed the process with this modification. The Goldstone boson
  scattering amplitudes are very good approximations to the scattering
  of longitudinal EW gauge bosons by means of the Goldstone boson
  equivalence theorem. By projecting the partial waves into their spin
  and isospin components, the optical theorem is used to determine the
  condition for perturbative unitarity in the same way as
  in~\cite{Lee:1977yc}. Further details are listed in
  \cite{Kilian:2014zja}. At high invariant mass $M_{VV}$  of the
  $WW$-scattering system, the enhancement the crossection by
  $\frac{M_{WW}^8}{m_H^8}$ in comparison to the SM due to the dimension
  eight operators are dominant. The coefficients of the higher-dimensional
  operators are chosen within current LHC bounds. We concentrated to the
  like-sign $WW$ scattering as this is the cleanest channel at the LHC
  with the smallest backgrounds. It only appears in the isospin two
  channel. In the light red band, we plotted the unitarity limit by
  demanding that the isospin partial waves $\mathcal{A}_{20}$ and $\mathcal{A}_{22}$ for
  isospin two and spin zero and two, respectively, are saturated,
  i.e. reaching their maximally allowed value of $32\pi$.

  The predictions of the dimension eight operators violate the unitarity limit and
  become unphysical in an energy regime, which can be tested at the LHC.
  Naively, one could introduce a cut-off to forbid these unphysical
  events manually (a prescription also partially used by ATLAS and CMS,
  known as 'event clipping'). Such a cutoff could also be motivated
  theoretically by the argument that these events could have never
  arisen in a UV-complete theory. However, this leads to a sharp edge in
  the distribution (at level of the vector bosons) which does not
  resemble any sensible approximation to a UV-complete theory, and
  furthermore there are also experimental restrictions for doing so:
  In case of the $W^+W^+$ scattering, the final state includes two
  neutrinos and the $WW$ invariant mass cannot be experimentally
  reconstructed. Other methods to treat this high-energy regime are by
  means of so-called form factors which, however, depend on at least two
  parameters, the exponent of the momentum dependence in the
  denominator (the 'multipole' parameter) and the cutoff scale which a
  priori has nothing to do with the scale $\Lambda$ appearing in front
  of the Wilson coefficients.

  \begin{figure}[bt]
    \begin{center}
      \includegraphics[width=0.8 \textwidth]{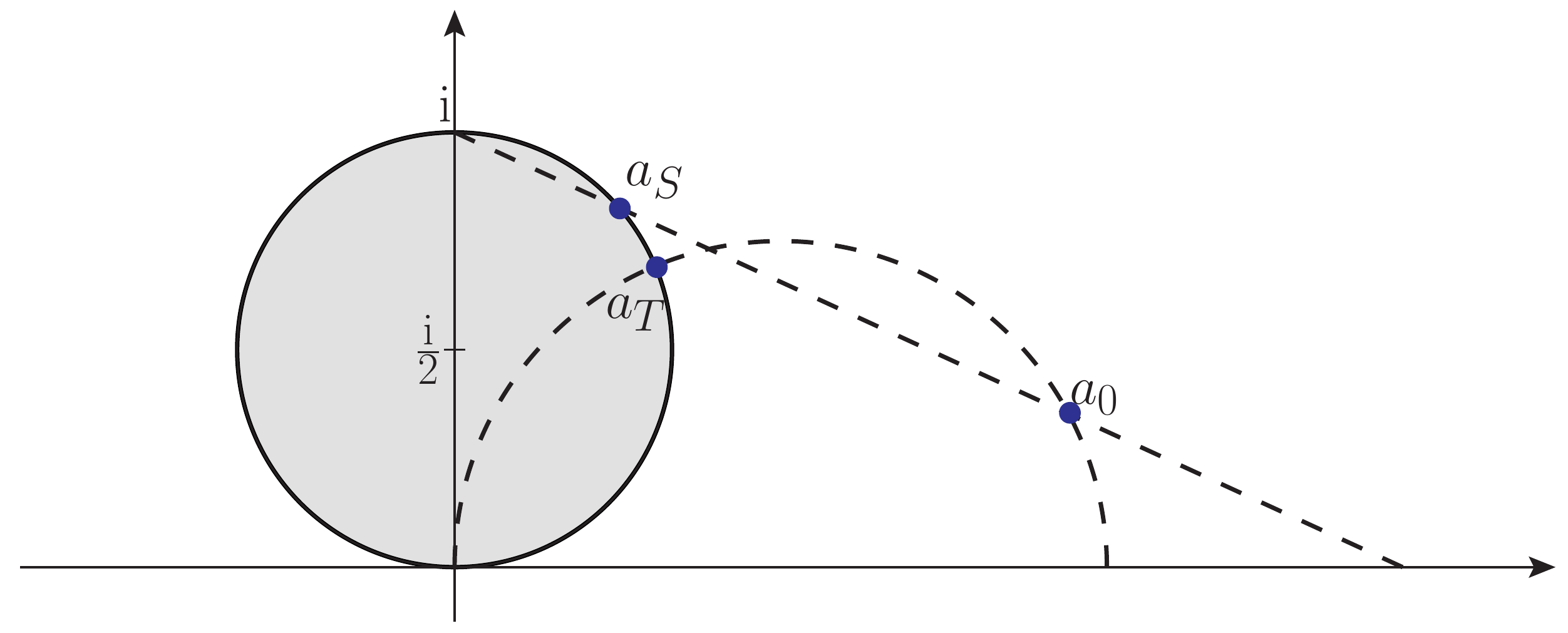}
    \end{center}
    \caption{Geometrical representation: stereographic projection vs
      Thales projection.}
    \label{fig:ThalesStereo2}
  \end{figure}

  In order to have a meaningful description that does not depend on any
  additional parameters, we introduce the $T$-matrix
  unitarization scheme (cf.~Fig.~\ref{fig:ThalesStereo2}
  in terms of the Argand circle, as described below) as a general
  extension of the $K$-matrix unitarization to provide event samples,
  which satisfy the unitarity bound. The T-matrix scheme
  is applicable for cases where the amplitude has an imaginary
  part itself already, and is also defined without relying on a
  perturbative expansion. For more details cf.~\cite{Kilian:2014zja}. The
  right-hand side of Fig.~\ref{fig:ww_unit} shows the damping of the
  cross sections for high energies due to the saturation of the
  amplitudes.

  In the first measurement of anomalous quartic gauge couplings in vector boson
  scattering by ATLAS \cite{Aad:2014zda}, the non-linear representation as
  defined in \cite{Alboteanu:2008my} is used to probe anomalous quartic gauge
  couplings $\alpha_4$ and $\alpha_5$,
  \begin{alignat}{3}
    \label{LL-alpha4}
    \LL_{\alpha_4}&=
    & &\alpha_4\ &&
    \tr{  \vV_\mu \vV_\nu }
    \cdot \tr{  {\vV^\mu} \vV^\nu } \, ,
    \\
    \label{LL-alpha5}
    \LL_{\alpha_5}&=
    & &{\alpha_5}\ &&
    \tr{ \vV_\mu \vV^\mu }
    \cdot \tr{  \vV_\nu \vV^\nu } \, .
  \end{alignat}

  $\mathcal{L}_{\alpha_4}$ and $\mathcal{L}_{\alpha_5}$ include the subset of couplings
  of $\mathcal{L}_{S,0}$ and $\mathcal{L}_{S,1}$ excluding all anomalous couplings
  with at least one Higgs. However,
  anomalous couplings involving the Higgs can be neglected
  in a study of vector boson scattering at the LHC due to the small Yukawa
  interaction of incoming quarks and specialized triggers for outgoing vector bosons.
  Therefore, both parametrization are equivalent to study anomalous quartic gauge
  couplings in vector boson scattering processes.
  Limits on the anomalous couplings $\alpha_4$ and $\alpha_5$
  can be directly translated to limits of $F_{S,0}$ and $F_{S,1}$, with or without
  K/T-matrix unitarization, via
  \begin{subequations}
    \begin{align}
      F_{S,0} &= 16 \frac{\alpha_4}{v^4} \, ,\\
      F_{S,1} &= 16 \frac{\alpha_5}{v^4} \, .
    \end{align}
    \label{eq:conversion}
  \end{subequations}
  For example, a conversion of the observed one-dimensional $95 \%$ interval
  in \cite{Aad:2014zda}  leads with eq. \eqref{eq:conversion} to following observed
  $95 \%$ interval for $F_{S,0}$ and $F_{S,1}$:
  \begin{subequations}
    \begin{align}
      -590 \;\mathrm{TeV}^{-4}<& F_{S,0} < 680 \;\mathrm{TeV}^{-4}
        \quad \mathrm {for} \quad F_{S,1}=0 \, ,  \\
      -420 \;\mathrm{TeV}^{-4}<& F_{S,1} < 510 \;\mathrm{TeV}^{-4}
        \quad \mathrm {for} \quad F_{S,0}=0 \, .
    \end{align}
  \end{subequations}

\begin{figure}[tb]
    \begin{center}
      \subfloat[SM]{\includegraphics[scale=0.32]{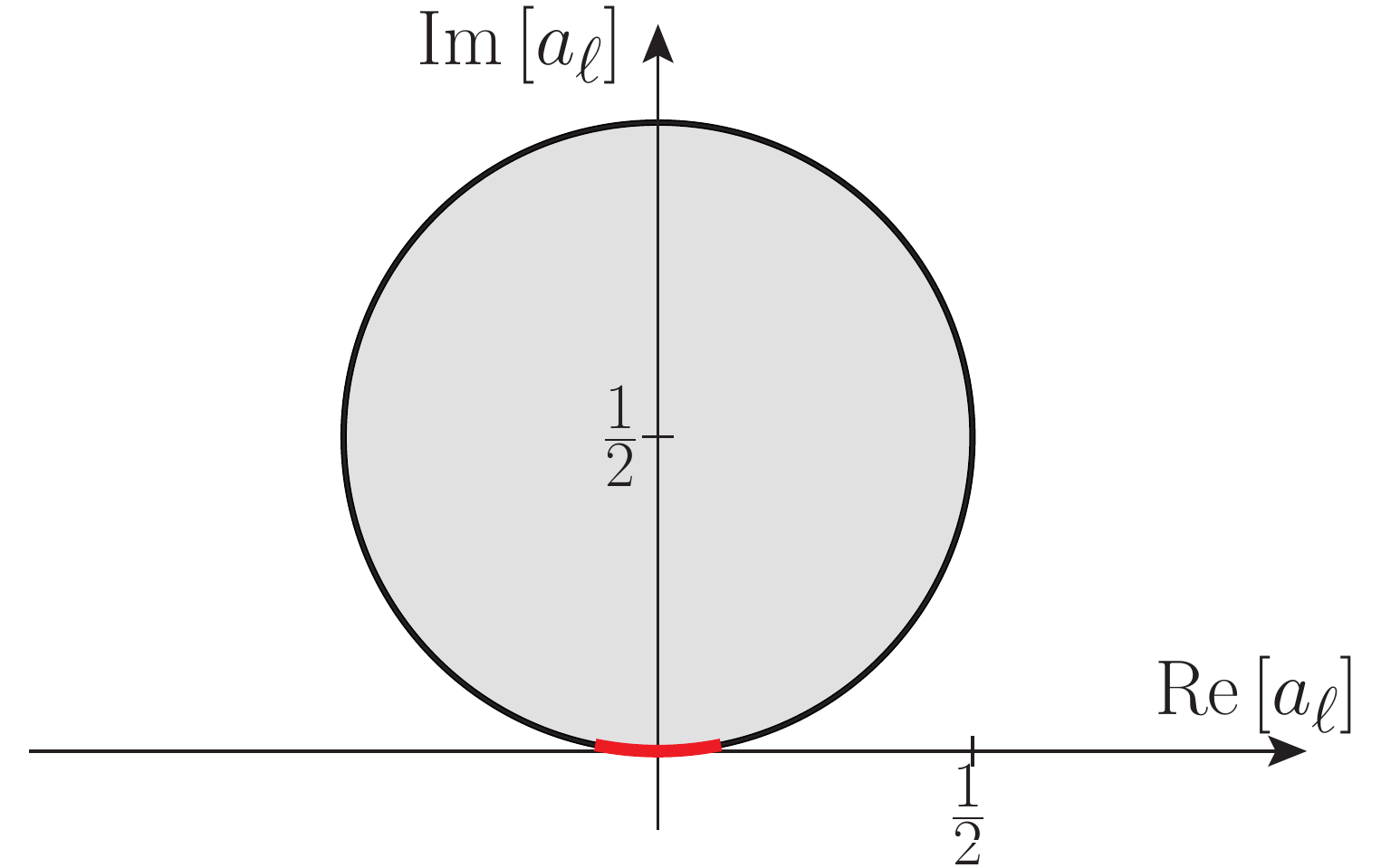}
        \label{subfig:sm}}
      \subfloat[Bare EFT, high-energy]{\includegraphics[scale=0.32]{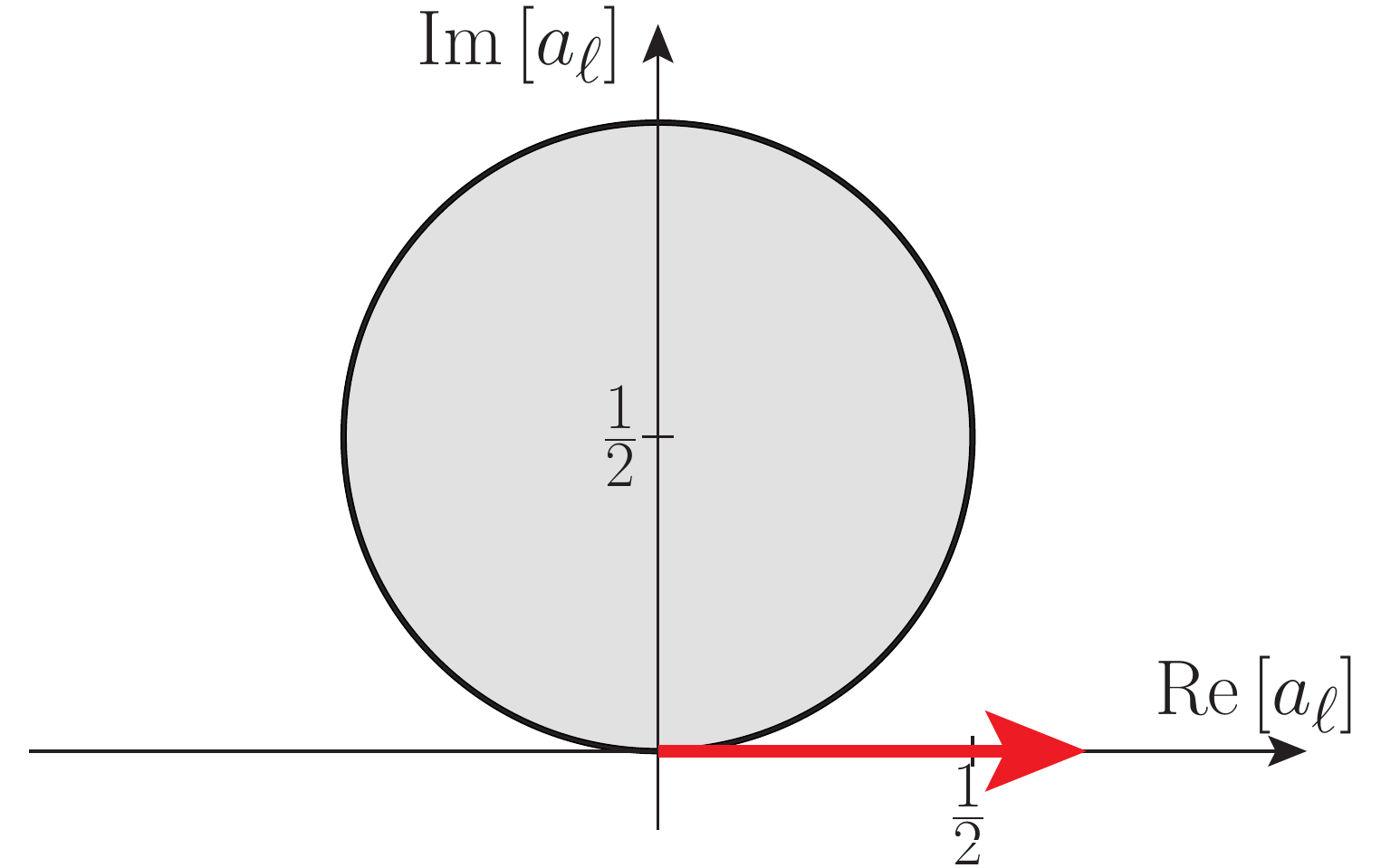}
        \label{subfig:eft}}\\
      \subfloat[Saturation]{\includegraphics[scale=0.32]{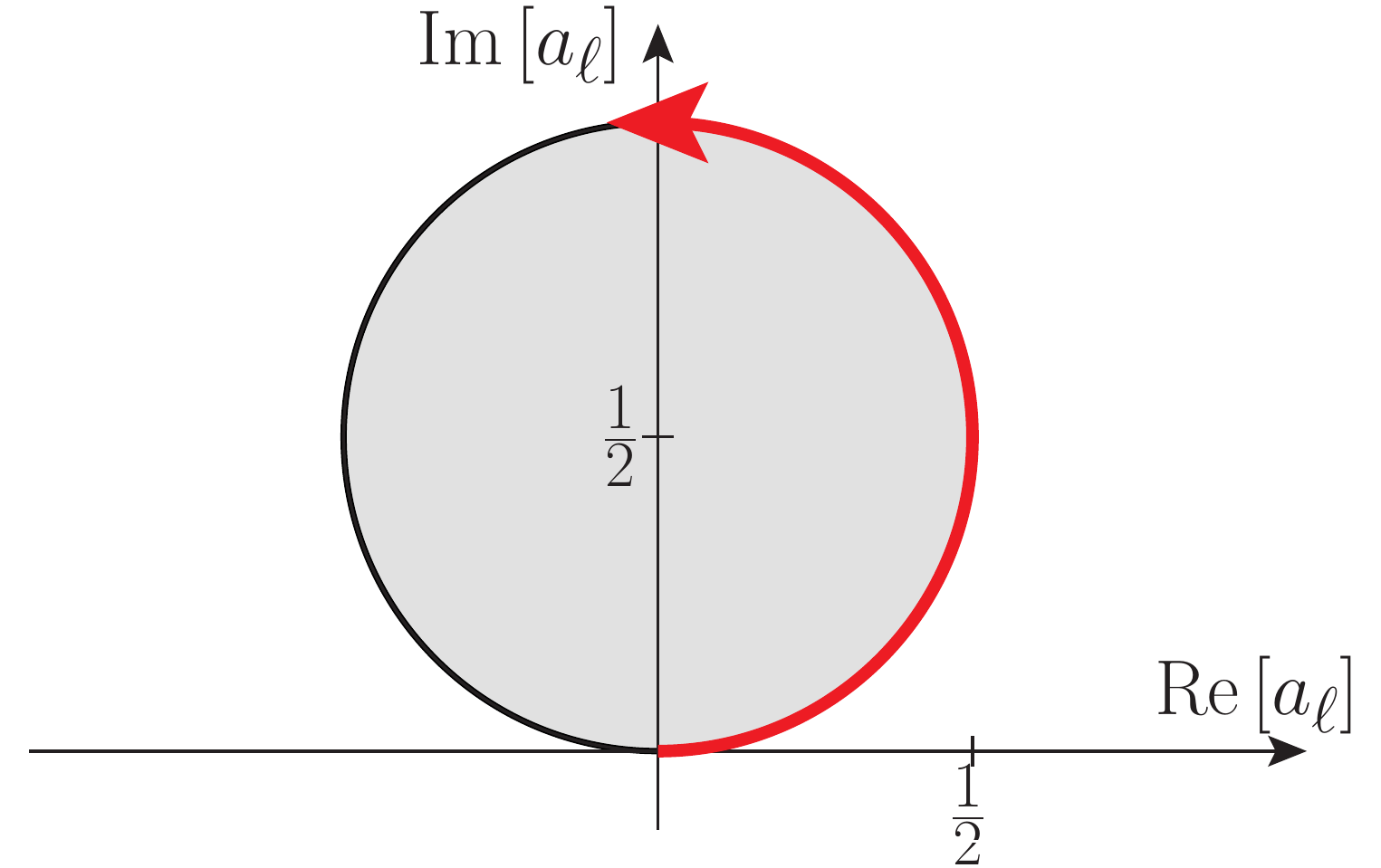}
        \label{subfig:saturation}}
        \subfloat[Inelastic channels]{\includegraphics[scale=0.32]{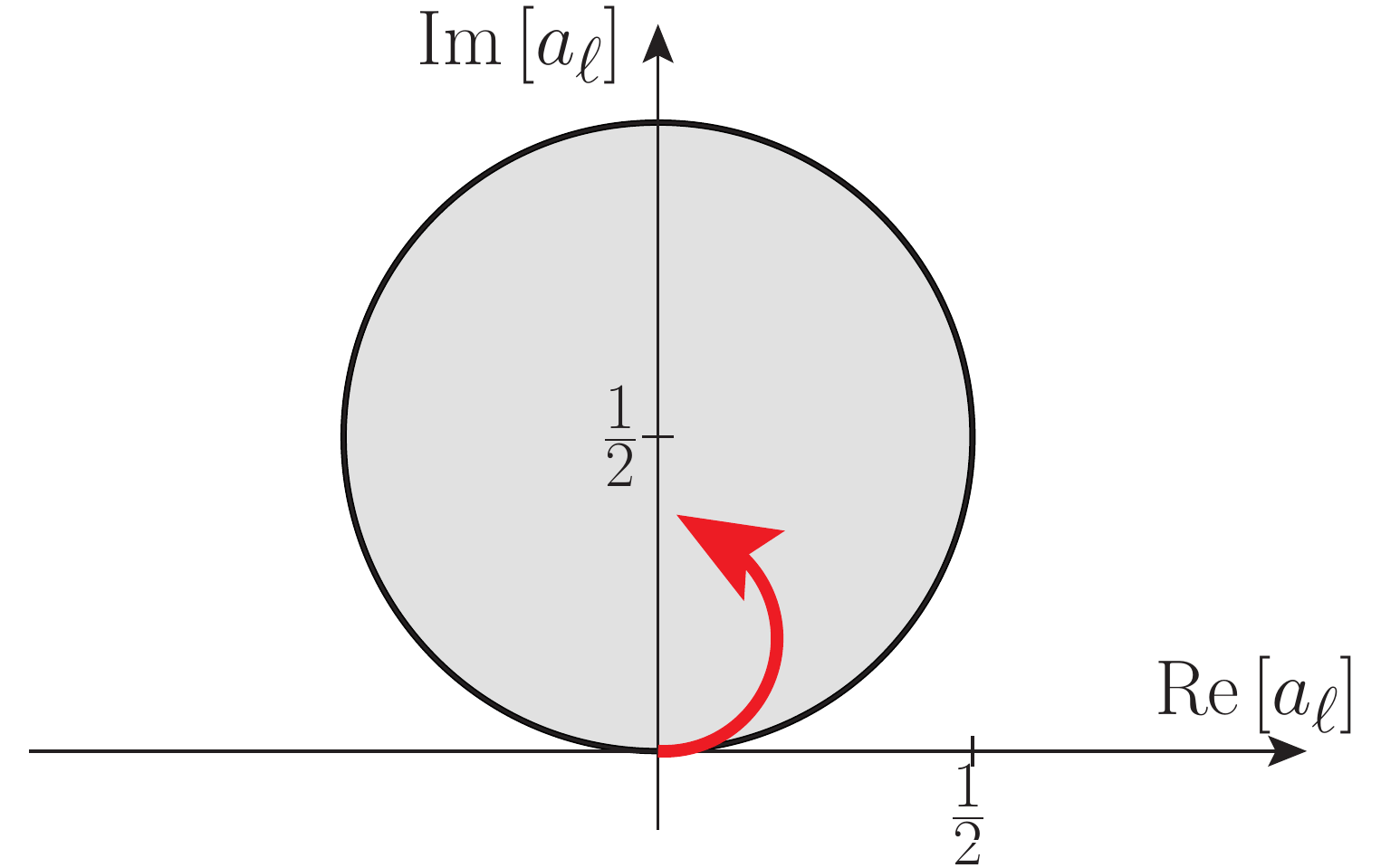}
          \label{subfig:inelastic}}
        \subfloat[Resonance]{\includegraphics[scale=0.32]{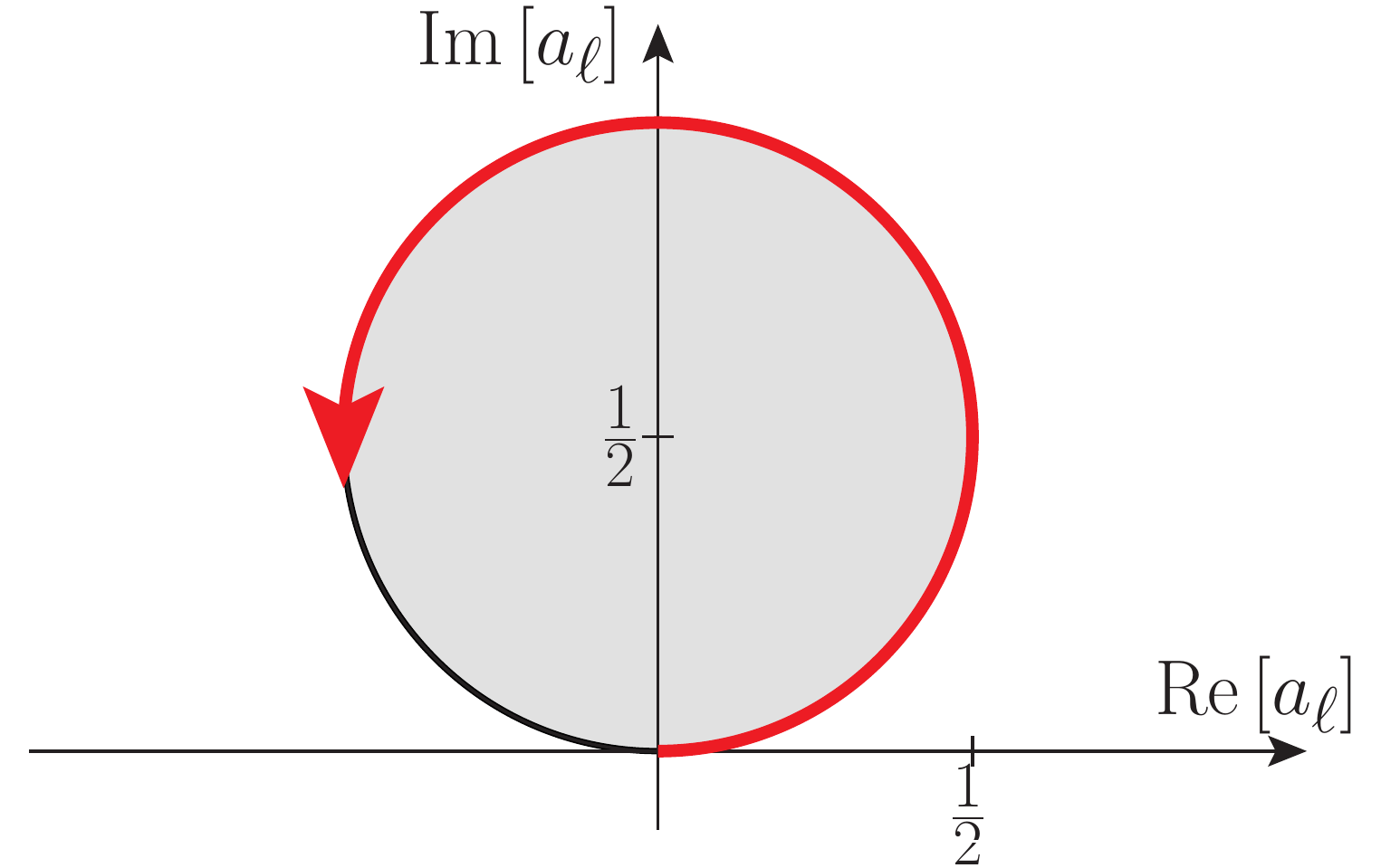}
          \label{subfig:resonance}}
    \end{center}
    \caption{Possible scenarios for scattering amplitudes respecting the Argand circle.}
    \label{fig:argand}
  \end{figure}

  The T-matrix scheme is only one possible extrapolation
  for high-energy scenarios. All physical scenarios have to fullfil the
  unitarity condition which is graphically represented by the Argand
  circle. If no new physics is involved in the electroweak sector, the
  elastic scattering amplitude of the Standard model will
  stay at the origin on the bottom of the Argand circle
  (Fig.~\ref{subfig:sm}). If the EFT is naively added, amplitudes start
  to rise and will leave the Argand circle to finally
  violate unitarity (cf.~Fig.~\ref{subfig:eft}, as there are no new degrees
  of freedom in the strict EFT, the amplitude can never develop an
  imaginary part to return to the Argand circle). To remedy this
  unphysical behavior of the amplitude, unitarization prescriptions are
  introduced to project the amplitude back onto the Argand circle. T-matrix
  unitarization saturizes the amplitude, in the sense that it is equivalent
  to an infinitely broad resonance at infinity, similarly to a strongly
  interacting continuum present over an extended range in momentum
  space. Practically, it will project the corresponding isospin-spin amplitude to its
  maximally allowed absolute value at high energies.
  Another option to correct the unphysical EFT prediction is
  using the form-factor scheme, a possible case of entering the
  inelastic regime with additional channels opening up
  (cf.~Fig.~\ref{subfig:inelastic}).  A third approach would be the
  addition of explicit resonances (either weakly or strongly coupled), which
  could be (part of the) origin for the dim-8 operators
  (cf.~Fig.~\ref{subfig:resonance}).  Here, the amplitude will ideally fall
  again beyond the resonance, but could show a rise again due to
  continuum contributions or the onset of a further resonance.

  \section{Resonances and Simplified Models}

  As the LHC is intended to be a discovery machine, it might be
  advantageous to assume that a new resonance or particle might be
  within the kinematic reach of the machine, especially given the high
  amount of luminosity to be collected in runs II and III. In order to
  be as general as possible in studying what kind of resonances could
  show up in vector boson scattering -- specific models would be Two-Higgs
  double models, including the (N)MSSM, Composite Higgs, Little
  Higgs~(for limits cf.~e.g.~\cite{LittleHiggs}), Twin Higgs, etc. --
  we classify all resonances that can couple to the electroweak diboson
  systems according to their spin and isospin quantum numbers. For
  simplicity, we neglect couplings to photons, but of course they are
  present due to EW gauge invariance. These possible resonances can be
  categorized in terms of the approximate $SU(2)_L \times SU(2)_R$ symmetry,
  which is a good approximation for weak boson scattering, and the
  spin. The $(0,0)$ and the $(1,1)$ representations of the $SU(2)_L
  \times SU(2)_R$ are abbreviated as isoscalar and isotensor,
  respectively. We can distinguish the resonances for elastic vector
  boson scattering into an isoscalar scalar $\sigma$, an isoscalar tensor
  $f$, an isotensor scalar $\phi$ and an isotensor tensor $X$. The
  interaction with longitudinal vector bosons is modeled by the
  following currents:
  \begin{subequations}
    \begin{alignat}{2}
      J_{\sigma} &= F_\sigma&&
      \tr{ \left ( \vD_\mu \vH \right )^\dagger \vD^\mu \vH} \, ,\\
      J_{\phi} &= F_\phi&&
      \left (
      \left ( \vD_\mu \vH \right )^\dagger \otimes \vD^\mu \vH
      +\frac {1}{8} \tr{\left ( \vD_\mu \vH \right )^\dagger \vD^\mu \vH }
      \right )\tau^{aa} \, , \\
      J^{\mu \nu}_f&=
      F_f &&\left (
      \tr{ \left ( \vD^\mu \vH \right )^\dagger \vD^\nu \vH}
      - \frac{c_f}{4} g^{\mu \nu}
      \tr{ \left ( \vD_\rho \vH \right )^\dagger \vD^\rho \vH}
      \right)  \, , \\
      J^{\mu \nu}_{X }&=
      F_X& & \Bigg [
        \frac{1}{2} \left (
        \left ( \vD^\mu \vH \right )^\dagger \otimes \vD^\nu \vH
        + \left (  \vD^\nu \vH \right )^\dagger \otimes \vD^\mu \vH
        \right )
        - \frac{c_X}{4} g^{\mu \nu}
        \left ( \vD_\rho \vH \right )^\dagger \otimes \vD^\rho \vH \notag \\
        &&&
        +\frac{1}{8} \left (
        \tr{\left ( \vD^\mu \vH \right )^\dagger \vD^\nu \vH}
        -  \frac{c_X}{4} g^{\mu \nu}
        \tr{\left ( \vD_\rho \vH \right )^\dagger \vD^\rho \vH}
        \right )
        \Bigg ] \tau^{aa} \, .
    \end{alignat}
  \end{subequations}
  Here, $\vH = \frac12 \left(\mathds{1} (v+H) -i w^a \tau^a\right)$, and
  $\tau^{aa}$ is the tensor-product representation for the isotensor
  case. With those resonances at hand, parameterized simply by their
  masses and widths, together with the currents above, one can integrate
  them out again and derive the corresponding Wilson coefficients of the
  dim-8 operators $\mathcal{O}_{S,0}$ and $\mathcal{O}_{S,1}$ in the
  section before, for all cases considered above. The coefficients
  are listed in table \ref{table:gamma}.

  \begin{figure}[tb]
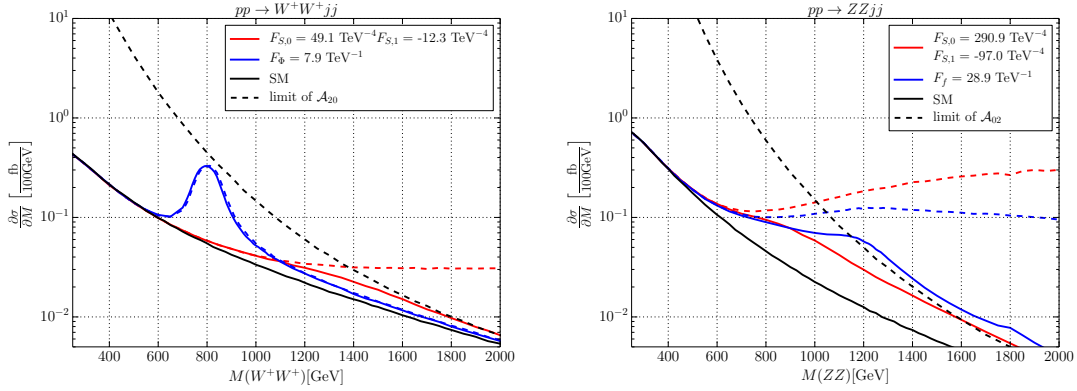

    \centering
    \includegraphics[width=0.48\linewidth]{{{ppWpWp_p_800-0_1dat_pp2}}}
    \includegraphics[width=0.48\linewidth]{{{ppZZ_f_1200-0.4dat_pp2}}}
    \caption{
      \label{fig:isoscalar-tensor}
      Differential cross sections of a scalar-isotensor resonance (left)
      and an isoscalar-tensor resonance (right). Solid line: unitarized
      results, dashed lines: naive result,
      black dashed line: limit of saturation of $\mathcal{A}_{22}$ $(W^+W^+)$ /
      $\mathcal{A}_{02}$ $(ZZ)$. Cuts: $M_{jj} > 500$ GeV;
      $\Delta\eta_{jj} > 2.4$; $p^j_T > 20$ GeV; $|\eta_j| > 4.5$. Left:
      $pp\rightarrow W^+W^+jj$, scalar-isotensor
      with $m_\phi= 800 \, \mathrm{GeV}$ and $\Gamma_\phi= 80 \,
      \mathrm{GeV}$, right: $pp\rightarrow ZZjj$, strongly interacting
      isotensor scalar with $m_f= 1200 \, \mathrm{GeV}$ and $\Gamma_f=
      480 \, \mathrm{GeV}$.}
  \end{figure}

  \begin{table}[tb]
    \begin{center}
      \begin{tabular}{c| M{2 cm}  M{2 cm}  M{2 cm} M{2 cm} N}
        & $\sigma$ & $\phi$  & $f$  &  $X$ & \\
        \hline
        $F_{S,0}$ & -- & $2$ & $15$& 5 &\\[4 ex]
        $F_{S,1}$ & $1/2$ & -$1/2$ & -$5$&  -35 &
      \end{tabular}
    \end{center}
  \caption{
  Relation of resonance width $\Gamma$ and mass $m$ to the corresponding dimension
  eight operator coefficients in the low-energy  effective field theory.  The
  factors listed in the table have to be multiplied by $32 \pi \Gamma/m^5$.}
  \label{table:gamma}
  \end{table}

  Tensor resonances as they could arise as Kaluza-Klein recurrences of a
  higher-dimensional gravity theory, but also as analogues to tensor
  mesons in a composite model, are particularly interesting. They
  usually give the largest signal contributions, as here the maximum
  number of spin components are involved in the scattering, namely five,
  compared to scalar and vector cases. There is a substantial difference
  in the theoretical treatment of those intrinsic spin degrees of freedom
  when dealing with the tensor resonance on-shell and off-shell. In a
  full Monte-Carlo simulation (cf. below), one actually simulates the
  final state and always has the tensor resonance in off-shell
  configurations. Using the analogue of unitarity gauge for tensors, the
  propagators lead to a bad high-energy behavior of the amplitudes.
  A symmetric tensor field $f_{\mu\nu}$ has 10 components which are
  reduced by the on-shell conditions to five physical components
  within the Fierz-Pauli Lagrangian~\cite{Fierz:1939ix}. These
  conditions are the tracelessness, $f_\mu^{\phantom{\mu}\mu} = 0$ and
  the transversality, $\partial_\mu f^{\mu\nu} = 0$. However,
  these Fierz-Pauli conditions are
  not valid off-shell, so we use the St\"uckelberg
  mechanism~\cite{Stueckelberg:1900zz,Stueckelberg:1941th} to
  make the off-shell high-energy behavior explicit. Onshell, there is only the
  tensor field, $f^{\mu\nu}$, while off-shell there is a vector field,
  $A^\mu \sim \partial_\nu f^{\mu\nu}$, which corresponds to the
  transversality condition, a scalar implementing the fully contracted
  transversality, $\phi \sim \partial_\mu \partial_\nu f^{\mu\nu}$, and
  another scalar corresponding to the tracelessness, $\sigma \sim
  f_\mu^{\phantom{\mu}\mu}$. By gauge fixing, one of the scalar degrees
  of freedom is redundant: $\sigma = - \phi$. The technical details
  together with the full Lagrangians and currents for the Fierz-Pauli as
  well as the St\"uckelberg picture can be found
  in~\cite{Kilian:2014zja}.

  Fig.~\ref{fig:isoscalar-tensor} shows two examples how differential
  invariant mass distributions of the diboson system behave at the LHC
  in the presence of such resonances. Both plots show different
  resonances in different scenarios: the left plot a narrow isotensor
  scalar with mass $m_\phi = 800$ GeV and width $\Gamma_f = 80$ GeV, the
  right one a strongly-interacting scenario with a broad
  isoscalar-tensor resonance of mass $m_f = 1.2$ TeV and width $\Gamma_f
  = 480$ GeV. The left plot shows the like-sign $W^+W^+$ channel, the
  right one the opposite-sign $W^+W^- \to ZZ$ channel,
  respectively. General cuts for selection and signal/background
  enhancement are shown in the caption. The full black line is the SM,
  the black dashed line shows the corresponding unitarity limit of the
  leading partial wave amplitude, the full blue line shows the SM with
  the corresponding resonance, while full red line depicts the
  approximation with the two Wilson coefficients, $F_{S,0}$ and
  $F_{S,1}$. Clearly, if explicit resonances are in the kinematic reach
  of the LHC, the EFT is no longer a viable approximation in any
  case. Note that even in the simulation with an explicit resonance,
  T-matrix unitarization has been applied to unitarize the high-energy
  tail of the distribution. As here the amplitudes do have explicit
  complex poles, T-matrix unitarization is actually needed.

\begin{figure}[tb]
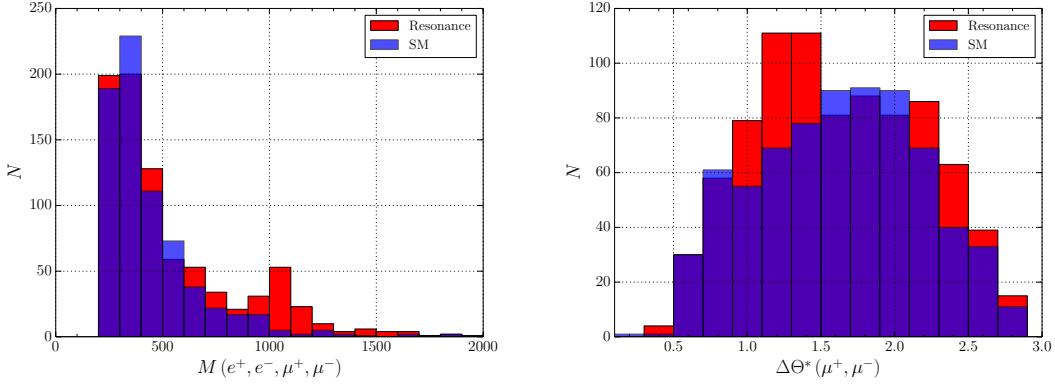

    \centering
    \includegraphics[width=0.48\linewidth]{{{VBS_ppe2mu2_f_r-0.1_1000-mww}}}
    \includegraphics[width=0.48\linewidth]{{{VBS_ppe2mu2_f_r-0.1_1000-theta_star}}}
    \caption{\label{fig:isoscalar_tensor_full}
      Isoscalar-tensor resonance at $m_f=1000$ GeV and
      $\Gamma_f $=100 GeV $pp\rightarrow e^+ e^- \mu^+ \mu^- jj$ at
      $\sqrt{s} = 14 \, \mathrm{TeV}$  with luminosity of $3000 \,
      \mathrm{fb}^{-1}$, with cuts $M_{jj} > 500$ GeV; $\Delta\eta_{jj} > 2.4$;
      $p^j_T > 20$ GeV; $|\eta_j| > 4.5$; $100\; \mathrm{GeV} >
      M_{e^+e^-} > 80\; \mathrm{GeV}$; $100\; \mathrm{GeV} > M_{\mu^+\mu^-} > 80\; \mathrm{GeV}$.
    }
  \end{figure}

  We have implemented the complete set of longitudinal dimension-6 and dimension-8
  operators together with the prescription
  of K-/T-matrix unitarization (for longitudinal VBS) in the Monte Carlo
  event generator
  \texttt{WHIZARD}~\cite{Kilian:2007gr,Moretti:2001zz}. It contains a
  quite elaborate machinery for QCD precision physics, where it uses the
  color flow formalism~\cite{Kilian:2012pz}, it has its own parton
  shower implementations~\cite{Kilian:2011ka}, and quite recently has
  successfully demonstrated its QCD NLO
  capabilities~\cite{WHIZARD_NLO}. \texttt{WHIZARD} has been used for a
  plethora of BSM studies, and is able to read in external models,
  e.g. via~\cite{Christensen:2010wz}. Using this implementation, we
  simulated vector boson scattering at the LHC with its design energy of
  $\sqrt{s} = 14$ TeV for all kinds of narrow and wide resonances of
  different spin and isospin. Fig.~\ref{fig:isoscalar_tensor_full} shows
  an example of an isoscalar tensor resonance of mass $m_f = 1$ TeV and a
  width of $\Gamma_f = 100$ GeV in the scattering of opposite-sign $W$s
  into two $Z$s. A standard set of selection cuts are mentioned in the
  caption of the figure. The left plot shows the invariant mass of the
  diboson system, which in this case is fully reconstructible, while the
  right plot shows the distribution of the opening angle of the two
  muons from one of the $Z$s. The latter is one of the angular
  observables that could be used to discriminate the spin of such
  resonances. More examples can be found in~\cite{Kilian:2014zja}.


  \section{Conclusions}

  The search for new physics in the electroweak sector in vector boson scattering at
  the LHC can be studied in the context of effective field theory, however, the
  introduction of dim-6 and dim-8 operators leads to a very limited range of
  applicability of the EFT ansatz. In many models, dim-8 operators could be the
  leading contributions where tree-level effects are forbidden by symmetries,
  and first contributions come in at the one-loop level, i.e. at dim-8. LHC as a
  hadron collider probes a vast range of energy scales, and high-energy events
  tend to (over-)dominate the exclusion limits (or search potentials) for new
  models. In most cases this is due to wrong assumptions on the underlying
  model, if EFT-based approaches in regimes are used where perturbative
  unitarity is lost. We studied examples of a dim-6 and two dim-8 operators and
  derived unitarity limits for the different spin and isospin channels in the
  scattering of longitudinal electroweak vector bosons. Then, a unitarization
  method, T-matrix unitarization, that is parameter-free and that is an
  extension of the "classic" K-matrix unitarization has been applied to produce
  results that are physically meaningful. The T-matrix unitarization has certain
  advantages, as it is defined for amplitudes that are intrinsically complex,
  and does not rely on the existence of a perturbative expansion. For weakly
  coupled amplitudes without imaginary parts it is identical to K-matrix
  unitarization. This procedure is not just an academic exercise, it allows to
  produce Monte Carlo events that could actually come from a quantum field
  theory realized in nature. Furthermore, it is itself a possible limit of a
  a strongly interacting continuum like in QCD or close to a
  quasi-conformal fixed point, or it could correspond to a strongly interacting
  model right below the onset of a new resonance that is just a little bit
  outside the kinematical reach of LHC. Even if the T-matrix prescription is not
  the correct extrapolation of the EFT, it can be interpreted as upper bound of
  the corresponding elastic channels, e.g. isospin-spin channels, due to its
  saturizing character. In combination with its property, that it leaves an
  interaction matrix invariant, which already satisfies unitarity, it can always
  used as ``fail-safe''-mechanism to generate event samples, which respect
  fundamental physical properties. We show examples of cross sections as well as
  kinematic and angular distributions to show the effects between "bare" EFT and
  unitarized simulations.

  Beyond this parameter-less approach to new physics in vector boson scattering,
  we provided a set of simplified models taking the SM added by all possible
  resonances in the spin-isospin channels to which two EW vector bosons can
  couple. We focused on scalar and tensor resonances, while vector resonances
  are more complicated due to their potential mixing with the EW bosons. To
  account for effects of particularly strongly interacting models, in addition
  higher-dimensional operators can be added. Also, adding just single resonances
  does not lead to renormalizable models with sound high-energy behavior, hence,
  we also applied T-matrix unitarization to the simplified models. In order to
  start with a prescription that already has the best possible high-energy
  behavior, we isolated the scalar and vector degrees of freedom in massive
  tensor fields via the St\"uckelberg mechanism to represent explicitly the bad
  behavior of tensor propagators in unitarity gauge. We concluded with a fully
  differential example for a simplified model with an isoscalar tensor
  resonance.

  An updated study for vector boson scattering in future lepton
  colliders~\cite{Baer:2013cma,Behnke:2013lya} is completed in
  \cite{Fleper:2016frz}. Further work will be devoted to the study of transverse
  $W$ and $Z$ polarizations, the discussion of vector resonances as well a
  implementation of the T-matrix for $2\to3$ processes,
  which is relevant for triple weak boson production.

  \section*{Acknowledgements}
  MS thanks the organizers for a very enjoyable and interesting conference.

  \bibliographystyle{unsrt}

\end{document}